\documentclass[lettersize,journal]{IEEEtran}
\usepackage{amsmath,amsfonts}
\usepackage{algorithmic}
\usepackage{algorithm}
\usepackage{array}
\usepackage[caption=false,font=normalsize,labelfont=sf,textfont=sf]{subfig}
\usepackage{textcomp}
\usepackage{stfloats}
\usepackage{url}
\usepackage{verbatim}
\usepackage{graphicx}
\usepackage{cite}
\usepackage[range-units=single]{siunitx}
\usepackage[pdftex]{hyperref}
\begin{document}
\bstctlcite{IEEEexample:BSTcontrol}

\title{Controlled Shifts of X-ray Emission Lines Measured with Transition Edge Sensors at the Advanced Photon Source}

\author{Tejas Guruswamy, Orlando Quaranta, Lisa Gades, Umeshkumar Patel, and Antonino Miceli%
\thanks{This research was funded by Argonne National Laboratory LDRD proposals 2018-002 and 2021-0059; is supported by the Accelerator and Detector R\&D program in Basic Energy Sciences' Scientific User Facilities (SUF) Division at the Department of Energy; used resources of the Advanced Photon Source and Center for Nanoscale Materials, U.S. Department of Energy (DOE) Office of Science User Facilities operated for the U.S. DOE, Office of Basic Energy Sciences under Contract No. DE-AC02-06CH11357. \textit{(Corresponding author: Tejas Guruswamy)}}%
\thanks{Tejas Guruswamy (e-mail: tguruswamy@anl.gov), Orlando Quaranta, Lisa Gades, Umeshkumar Patel, and Antonino Miceli are with Argonne National Laboratory, Lemont, IL 60439 USA.}%
\thanks{Color versions of one or more of the figures in this article are available online at \protect\url{http://ieeexplore.ieee.org}.}%
}

\markboth{ASC2024-2EPo1D-05}%
{ASC2024-2EPo1D-05}

\maketitle

\begin{abstract}
The measurement of shifts in the energy of X-ray emission lines is important for understanding the electronic structure and physical properties of materials. In this study, we demonstrate a method using a synchrotron source to introduce controlled eV-scale shifts of a narrow line in between fixed-energy fluorescence lines. We use this to characterize the ability of a hard X-ray superconducting Transition Edge Sensor (TES) array to measure line shifts. Fixed fluorescence lines excited by higher harmonics of the monochromatic X-ray beam are used for online energy calibration, while elastic scattering from the primary harmonic acts as the variable energy emission line under study. We use this method to demonstrate the ability to track shifts in the energy of the elastic scattering line of magnitude smaller than the TES energy resolution, and find we are ultimately limited by our calibration procedure. The method can be applied over a wide X-ray energy range and provides a robust approach for the characterization of the ability of high-resolution detectors to detect X-ray emission line shifts, and the quantitative comparison of energy calibration procedures.
\end{abstract}


\section{Introduction}
\IEEEPARstart{X}{-ray}
fluorescence spectroscopy has been widely used as a powerful technique for studying the elemental composition and chemical structure of materials. It utilizes the characteristic X-ray emission lines produced by materials when they are excited with high energy X-rays. The absolute energy, relative amplitude, and detailed structure of these emission lines all provide valuable information about the electronic states and energy levels of the atoms in the material. 

One interesting phenomenon that can occur due to changes in chemical composition or physical state is the shift in the position and/or shape of these characteristic emission lines. A scientific application of this is in thermographic phosphors. These are luminescent materials that can be employed as temperature sensors by measuring the spectral or temporal changes in their emission under ultraviolet (UV) or X-ray excitation. While the temperature-dependent spectral response of thermographic phosphors under UV excitation has been well studied, the use of X-ray excitation for temperature measurements in these materials has not been as extensively explored~\cite{Westphal2021,Westphal2021a}. High energy X-rays have the unique advantage of being able to penetrate optically thick environments, making them suitable for temperature measurements in situations where UV excitation is challenging or impossible. 

In recent years, Transition Edge Sensors (TESs) have emerged as an advanced technology for high-resolution X-ray spectroscopy at synchrotron facilities~\cite{Kothalawala2024,Patel2022,Guruswamy2020,Guruswamy2021,Lee2019}. TESs are superconducting devices that operate in the vicinity of their superconducting transition temperature, where small changes in energy deposition give rise to large changes in electrical resistance. This sensitivity makes TESs particularly suitable for measuring subtle energy shifts in X-ray emission lines.

In this work, we describe a synchrotron experimental technique that provides a repeatable, fine-tunable line shift of a single line amongst other fixed lines, useful for the accurate characterization of the absolute energy measurement capabilities of X-ray detectors. This technique is particularly useful for TES spectrometers, which are able to achieve an energy resolving power $E/\Delta{}E \sim 1,000 - 10,000$. This means that eV or sub-eV control over the line shifts at hard X-ray energies is necessary, all without disturbing the fixed reference lines used for online calibration. In our demonstration experiment, after testing our TES array, we successfully complete the measurement of relative line shifts of a few eV but conclude that our ability to measure absolute line shifts is limited by our calibration accuracy.

\section{The Instrument}
\begin{figure}[htb]
\centering
\includegraphics[width=3.45in]{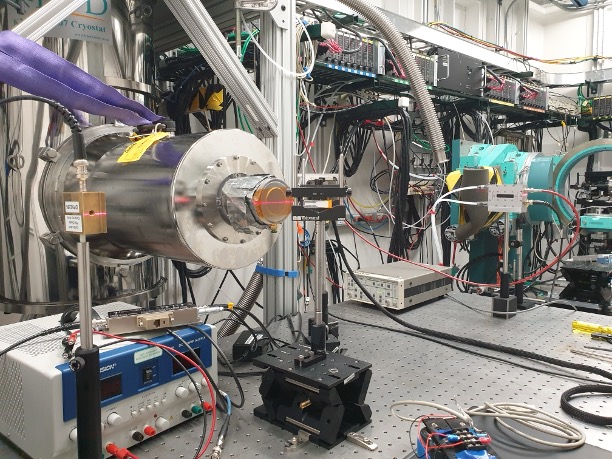}
\caption{Photo of the experimental setup at APS 1-BM-C beamline, with the Transition Edge Sensor instrument in the large cryostat towards the left. In this view, the X-ray beam travels from right to left. Also visible are an upstream ion chamber and a downstream PIN diode, which measure transmitted flux. The white cuvette at the center of the photo contains the sample under study.}
\label{beamline}
\end{figure}

The spectrometer characterized in this work is built upon the foundation of a FormFactor adiabatic demagnetization refrigerator (ADR) Model 107 cryostat~\cite{formfactor-web}. To correctly orient and bring the TES sensor closer to the sample under examination, the sensor itself is in a snout structure protruding out from the cryostat body. The instrument is installed at the Advanced Photon Source hard X-ray synchrotron beamline hutch 1-BM-C, in a 90\textdegree{} configuration with respect to the incident X-ray beam and the sample. The sample is oriented at 45\textdegree{} with respect to the beam and the sensor to maximize the illuminated area visible to the TES array. Fig. \ref{beamline} illustrates the typical experimental setup at 1-BM-C. A representative sample (here, a white powder in a cuvette) is on a stage in the middle of the picture. The red laser line indicates the expected height of the X-ray beam, which travels from right to left in the photo. To monitor the photon flux, the beam passes through multiple ion chambers before the sample, and terminates in a PIN diode.

When an X-ray photon hits the sample under study, it can induce the emission of secondary X-ray photons via several mechanisms, including elastic scattering, inelastic scattering, and fluorescence. Fluorescence only occurs when the incident photon energy is above one of the sample material's absorption edges, and causes the sample to re-emit photons at energies characteristic of the material. The sensor location at a 90\textdegree{} scattering angle maximizes the ratio of fluorescence photons to elastically scattered photons, as the incoming X-ray beam is horizontally polarized.

The sensor array used for this experiment was composed of 24 identical TESs fabricated at Argonne National Laboratory, all using a sidecar design. In this arrangement, the temperature-sensing superconducting bilayer and the X-ray absorber sit side-by-side on a suspended membrane, connected via a metal patch. The TESs are optimized for hard X-ray operation with \SI{780}{\micro{}m}-side square absorbers plated with \SI{3}{\micro{}m} of Au and \SI{17}{\micro{}m} of Bi.
We measure an average critical temperature $T_c \sim \SI{80}{mK}$, thermal conductance to the bath $G \sim \SI{360}{pW/K}$ and pulse characteristic decay time $\tau \sim \SI{5}{ms}$. Due to limitations of the available microwave frequency-multiplexed readout circuitry, only 12 pixels were actually read out in this experiment.

All TES data analysis was performed using the Microcalorimeter Analysis Software System (MASS)~\cite{mass_github}. It uses an optimal-filter based approach and includes post-filter corrections for some known systematic errors: baseline gain dependence, sub-sample arrival time dependence, and gain drift versus time~\cite{Fowler2016}. Each pixel is filtered independently with an optimal filter calculated offline (after data collection), based only on the average pulse shape from the current dataset under analysis. After semi-automatically matching a specified set of elemental fluorescence lines to the most prominent peaks in the resulting spectrum, an exact cubic spline is used to interpolate between the specified calibration points in log-log space, and all calibrated pixel spectra are co-added. This programmatic approach provides repeatable results and allows for the automated processing of a large number of datasets.

Fluorescence emission lines from foils of pure Au were used to determine the detector energy resolution prior to commencing beamline measurements. As the observed detector resolution was clearly much larger than any intrinsic line widths, and detailed line shape information for Au L-lines was not readily available, Gaussian line fits were performed for the Au L$_\alpha$ line on both the individual pixel and co-added spectra. We found, for the 12 pixels used in the measurements, a co-added Gaussian FWHM of \SI{28.6}{eV} with a standard deviation of \SI{2.4}{eV} between pixels. Experimental characterizations of Au list an overall intrinsic line width of \SI{7.78 \pm 0.12}{eV} (Au L$_{\alpha{}1}$) and \SI{8.08 \pm 0.16}{eV} (Au L$_{\alpha{}2}$)~\cite{Amorim1988}, so this suggests our instrument achieves an intrinsic resolution (FWHM) of $\sim$ \SI{27}{eV} at \SI{10}{keV} when co-adding all pixels, with a best single-pixel resolution approaching \SI{20}{eV}. This coincides with previous measurements of similar detector designs~\cite{Guruswamy2021}.

\section{The Method}
Energy calibration of an X-ray spectrometer is typically performed using fluorescence emission lines, for example those emitted from a sample whose elemental makeup is known. Using the expected energy and shape of these lines allows one to define a calibration curve that links the amplitude of current pulses from the detector to the photon energy. Although this is a standard approach for X-ray spectroscopy, it introduces a limitation in that it naturally fixes the position of all emission lines used for calibration to their expected energies. If the objective is to measure small changes in the shape or the position of certain emission lines, a different solution is needed: calibration using a set of reference lines distinct from the lines under study.

One approach is offline calibration, where one creates a calibration curve beforehand using a reference measurement, and then uses that curve to calibrate the subsequent experiment's data. This is a standard approach used for many semiconductor-based detectors. Although this works well enough for detectors with moderate energy resolution, it is very challenging to do with high energy resolution such as TESs, where even small drifts in the detector responsivity over the duration of an experiment can cause variations in the calibration of magnitude similar to the detector energy resolution. Essentially, the validity of the offline calibration is likely to be too short in time to be useful. Moreover, TESs, due to the nature of a superconductor biased in the transition between the superconducting and normal state and the electrothermal feedback circuit in which they are embedded, are sensitive to any changes in their electromagnetic and thermal environment, and even hysteretic: dependent on the path through which the sensor was brought to the operating point. For example, nearby trapped magnetic flux can affect the $T_c$ as well as the shape of the normal-to-superconducting transition ($R$), which is a function of both the temperature ($T$) and the bias current ($I$). The derivatives of the $R(T, I)$ function with respect to $T$ $\left(\alpha = \tfrac{T}{R}\tfrac{dR}{dT}\right)$ and $I$ $\left(\beta = \tfrac{I}{R}\tfrac{dR}{dI}\right)$ are two parameters which feed directly into the photon response of a TES~\cite{Irwin2005}. With significant effort, offline calibration curves can still be created based on a careful and complete characterization of the instrument under all possible experimental conditions (bias, bath temperature, electromagnetic environment, etc.). These calibration curves can then be used in conjunction with a gain tracking system, to constantly adapt to the changes in the detector response~\cite{Cucchetti2024,Porter2016}. Although functional (and often necessary in space missions), this approach is complex and is not generally practical in a synchrotron environment where conditions change frequently.

A related approach is a quasi-online calibration, where one periodically recalibrates the sensors against known emission lines throughout the entire experiment by switching back and forth between the sample and the reference under the X-ray beam~\cite{Yan2022} during a single measurement. This approach requires specific sample-rotation hardware which may or may not be compatible with the experimental setup. Sample environments at beamlines can be very complex, involving the control of various physical parameters: temperature, pressure, humidity, etc. In this work, we instead developed a setup which provided full online calibration, making both the ``unknown energy'' line under study and a range of known lines available at every moment during the measurement.

We designed a technique that used the capabilities of our synchrotron beamline to quantify the ability of our TES spectrometer to detect small changes in the position of specific emission lines in a very robust and repeatable way, allowing us to explore the absolute energy accuracy of our online calibration techniques. To do so, we used a foil of pure Au as the source of the reference lines (specifically, the L-lines), while we used the elastic scattering of the incident beam from the same foil as the line under study. The diffracted beam from our beamline double crystal monochromator (Si-111 reflection) can be tuned to specific energies with extremely narrow bandwidth ($E/\Delta{}E \sim \num{e4}$) and can be controlled to the order of $< \SI{1}{eV}$ by using a motor to change the crystal scattering angle. This allowed us to extremely precisely control the position of the elastic scattering line without disturbing any of our reference lines, representing the ideal proxy for a line-shift detection experiment. 

\begin{figure}[!t]
\centering
\includegraphics[width=3.45in]{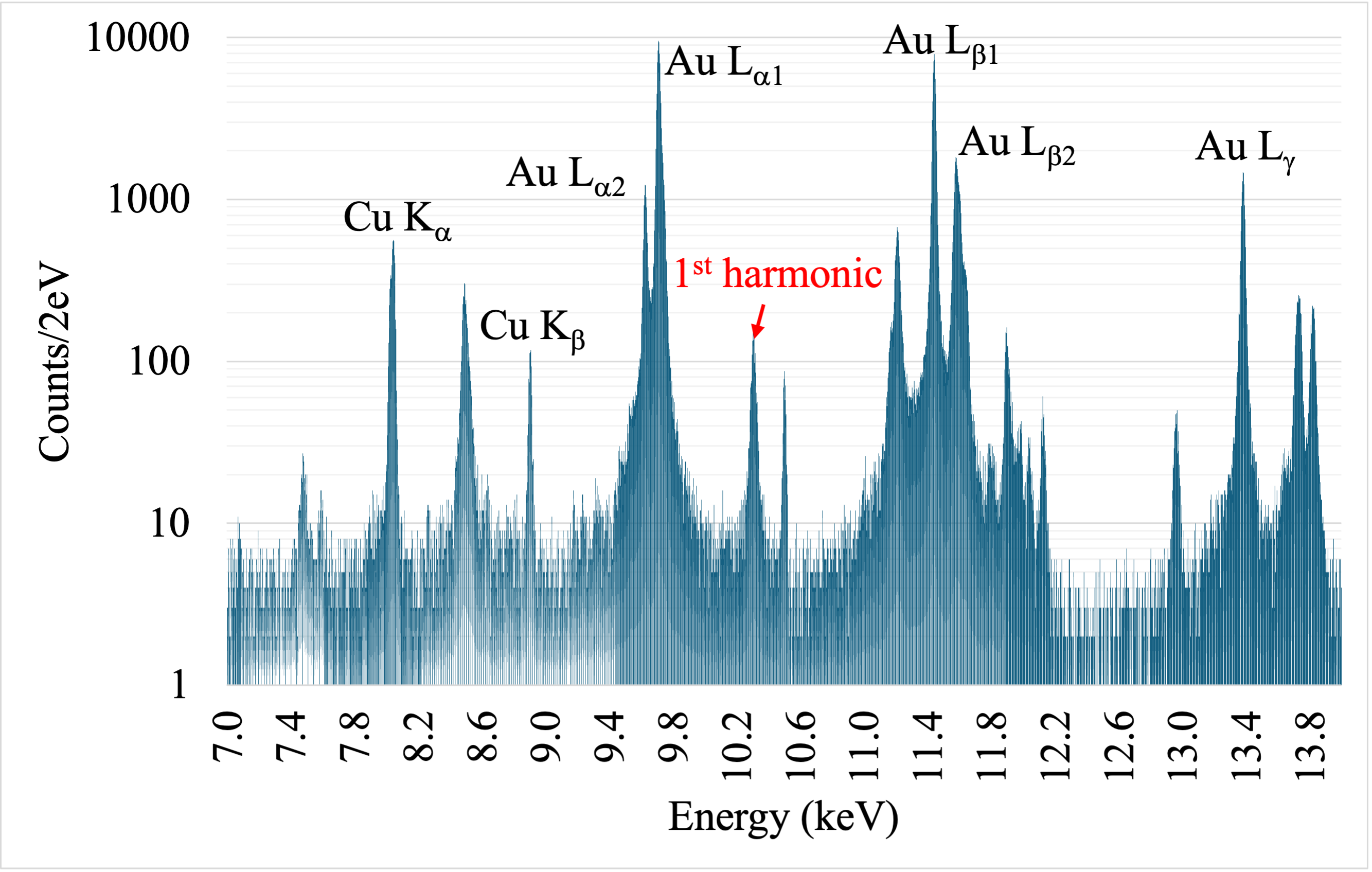}
\caption{Example co-added energy spectrum collected by the TES pixels (log scale). Labels indicate the elastic scattering line from the first harmonic of the incident X-ray beam, which is the line under study, and the various fluorescence lines, which are fixed in energy and used for calibration.}
\label{spectrum}
\end{figure}

As discussed in the previous section, our calibration scheme uses a cubic spline function to interpolate between specified calibration points in log-log space. To avoid extrapolation of the spline function beyond its well-constrained range, reference lines need to be available both below and above (in energy) the line under study. To excite the emission of fluorescence lines, the beam energy needs to be above the corresponding absorption edge; however, this would necessarily force it outside the range covered by those same calibration reference lines. In order to solve this apparent conflict, we used the higher-order reflections of our Si double crystal monochromator to excite fluorescence from the Au L lines, while still being able to have a variable-position line in the middle of the range from the primary reflection. The working principle of a monochromator is based on Bragg diffraction: a polychromatic beam (e.g. from the synchrotron bending magnet) hits one or two crystals at a specific angle, resulting in only a specific wavelength (and its harmonics) being reflected constructively. The harmonics have diminishing intensity with the order of the harmonic. In this experiment we used the primary and third harmonics, the closest accessible to the primary energy. By setting the first harmonic to lie between the Au L$_\alpha$ and L$_\beta$ lines, the third harmonic (with exactly three times that energy) is guaranteed to be sufficient to excite those same Au L lines. To equalize the intensity of the primary and third harmonics, thin metal foil filters which attenuate the first harmonic much more strongly than the third harmonic (Cu and Zn foils with absorption edges just below the first harmonic energy, each three attenuation lengths in thickness) were placed in the beam before the sample. A representative spectrum with the reference fluorescence lines and the elastic scattering line from the first harmonic is shown in Fig. \ref{spectrum}. The exact energy of the various fluorescence lines have been derived from Ref. \cite{nist_xraydb}.

In the spectrum, the various Au L emission lines are clearly visible, with the L$_{\alpha{}1}$ and L$_{\beta{}1}$ lines just below and above the first harmonic line respectively. More peaks are also visible from the Cu and Zn present in the upstream filters, and in the cryostat body near the sensor. After measuring each spectrum, the energy of the primary harmonic as measured by the TES was determined by fitting a Gaussian to the energy calibrated peak and taking the center value.

The experiment was repeated for varying incident beam energies, with a total range of around \SI{20}{eV} and centered on a nominal beam energy of \SI{10.5}{keV}. For each energy point the acquisition lasted 2 to 3 hours, to accumulate between \num{e3} and \num{e4} counts in the principal harmonic peak, while maintaining a realistic timeframe for the complete experiment with such a small array (12 pixels). This collection time could be significantly reduced by using a larger array. In principle, the error in the Gaussian fit of the primary harmonic peak should reduce as $1/\sqrt{N}$ with the number of counts in the peak, but as all measurements were histogrammed using the same bin size and all fits were performed over the same total energy range, we found no correlation in fit parameter uncertainty with total counts.

The variable energy line was always kept between the two anchor lines. It is important to note that there is some error between the nominal beam energy set with the monochromator controls and the actual beam energy. By measuring transmission through foils with known absorption edges, we can confirm that there was a mismatch between the nominally set energy and the actual beam energy obtained, caused by small shifts or drifts in the orientation of the monochromator crystal. This is likely due to a combination of heating effects on the monochromator caused by the incident high-intensity polychromatic beam, and offsets introduced in the intended versus actual position of the monochromator motor. A careful recalibration of the monochromator control system to correct these is possible but was not available to us at the time of our experiment. Therefore, the absolute value of the nominal beam energy we present is considered accurate only to within a few eV, but even with this systematic error we assessed that the proportionality of nominal vs actual beam energy remained sufficiently constant over the energy range needed for this experiment. 

\begin{figure}[!t]
\centering
\includegraphics[width=3.45in]{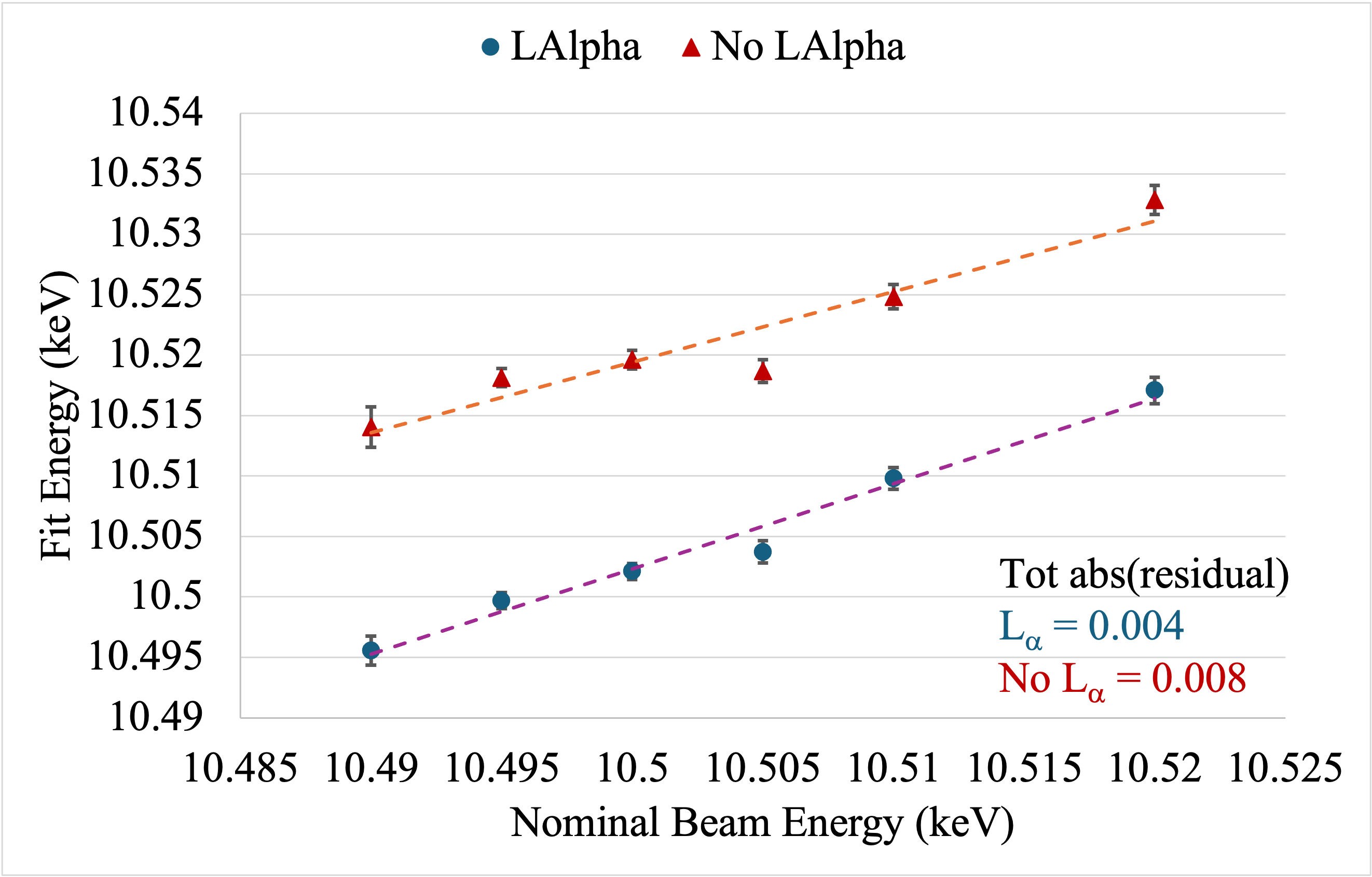}
\caption{Estimated energy from X-ray spectra vs nominal beam energy. In blue circles are represented measures using all the emission lines, while in red triangle the measures not using the L$_\alpha$ lines. In dashed lines are also represented the respective linear fits, together with total of the absolute value of the residuals. Error bars on each point represent only the uncertainty in the line fit parameters, and do not include any uncertainty introduced by the TES energy calibration.}
\label{shift}
\end{figure}

Fig. \ref{shift} illustrates the relation between the nominal primary harmonic energy and the energy as measured by fitting each TES-measured spectrum (with error bars from the fit only, not accounting for uncertainty from the calibration procedure). In one case (blue dots) both Au L$_\alpha$ and $L_\beta$ lines were used in the determination of the calibration function, while in the other (orange triangles) the Au L$_\alpha$ lines were deliberately ignored, meaning references only existed at energies above the line under study.

Both sets of data show a linear relationship with the nominal beam energy, indicating relative energy shifts are well captured by our current data analysis procedure. However, the dataset that used both sets of fluorescence lines in the calibration results in a significant shift of the estimated energies, moving them much closer to the nominal values. Moreover, the sum of the absolute value of the residuals when compared to a linear fit is a factor of two smaller. We also note that even in the improved calibration case, the scatter of the points does not appear to be completely accounted for by the fit parameter uncertainty (the error bars in the figure). The uncertainty introduced by the calibration procedure therefore appears to be significant and cannot be ignored. We conclude our current calibration approach on its own cannot achieve an absolute estimation of line energy with sub-eV precision, and that to achieve those accuracies, approaches like the ones described in Refs.~\cite{Yan2022} and \cite{Fowler2017} are needed.

Nonetheless, these results show the validity and simplicity of the overall technique, which allows one to measure the achievable absolute energy accuracy of an X-ray detector at arbitrary energies, and easily compare different variations of calibration strategies and combinations of reference lines available simultaneously with the line under study. In our case, our array detected line shifts in energy of a few eV even for TESs with moderate energy resolution ($> \SI{20}{eV}$). This experiment also reaffirms for us how fundamental it is for absolute energy measurement to have as many calibration lines as possible, as close to the region of interest as possible and on both sides of the region of interest, and to have a strong understanding of the uncertainty introduced by the calibration procedure. We note that MASS has recently added calibration uncertainty determination by Gaussian Process Regression~\cite{Fowler2022}, which will prove useful in future work along these lines.

\section{Conclusion}
In this work we described a technique to directly probe the ability of a Transition Edge Sensor array or other high-resolution spectrometer to detect X-ray fluorescence line shifts, based on the use of the both the primary and the third harmonic of a monochromatic X-ray beam at a synchrotron to enable a variable energy line as well as fixed fluorescence lines for online calibration. With this technique we showed our TESs, with a very basic online calibration strategy, can measure relative X-ray line shifts of the order of a few eV even when the best single-pixel detector resolution is only about \SI{20}{eV} at \SI{10}{keV}, proving their usefulness in the study of materials characterized by temperature and/or pressure dependent X-ray fluorescence~\cite{Westphal2021}. For absolute energy measurements or for smaller energy shifts, a more detailed calibration scheme is required.
Recently we undertook a series of experiments aimed at measuring an expected shift of \SI{5}{eV} or more in the position of fluorescence lines of a YAG:Dy thermographic phosphor in the temperature range \SI{25}{\celsius} to \SI{350}{\celsius} at APS using this TES spectrometer. In this experiment, a monochromatic beam was used to excite both the YAG:Dy sample, and a pure Zr foil, acting as calibration reference. The experiments didn't appear to show any indication of a shift in the Y K-lines with temperature. However, without a quantitative understanding of the type of line shifts our system could measure, it would not have been possible to rule out the presence of an effect. The technique described in this work allowed us to establish a technique to experimentally understand and verify our instrument's ability to measure such shifts, while comparing different calibration strategies. Moreover, this approach is extremely flexible, applicable over the wide range of energies attainable at a beamline.

\bibliography{line_shifts}

\begin{thebibliography}{10}
\providecommand{\url}[1]{#1}
\csname url@samestyle\endcsname
\providecommand{\newblock}{\relax}
\providecommand{\bibinfo}[2]{#2}
\providecommand{\BIBentrySTDinterwordspacing}{\spaceskip=0pt\relax}
\providecommand{\BIBentryALTinterwordstretchfactor}{4}
\providecommand{\BIBentryALTinterwordspacing}{\spaceskip=\fontdimen2\font plus
\BIBentryALTinterwordstretchfactor\fontdimen3\font minus
  \fontdimen4\font\relax}
\providecommand{\BIBforeignlanguage}[2]{{%
\expandafter\ifx\csname l@#1\endcsname\relax
\typeout{** WARNING: IEEEtran.bst: No hyphenation pattern has been}%
\typeout{** loaded for the language `#1'. Using the pattern for}%
\typeout{** the default language instead.}%
\else
\language=\csname l@#1\endcsname
\fi
#2}}
\providecommand{\BIBdecl}{\relax}
\BIBdecl

\bibitem{Westphal2021}
E.~R. Westphal, A.~D. Brown, E.~C. Quintana, A.~L. Kastengren, S.~F. Son, T.~R.
  Meyer, and K.~N.~G. Hoffmeister, ``Temperature-dependent {X-ray} fluorescent
  response from thermographic phosphors under {X-ray} excitation,''
  \emph{Applied Physics Letters}, vol. 119, no.~3, p. 034103, 2021.

\bibitem{Westphal2021a}
E.~R. Westphal, A.~D. Brown, E.~C. Quintana, A.~L. Kastengren, S.~F. Son, T.~R.
  Meyer, and K.~N.~G. Hoffmeister, ``Visible emission spectra of thermographic
  phosphors under {X-ray} excitation,'' \emph{Measurement Science and
  Technology}, vol.~32, no.~9, p. 094008, 2021.

\bibitem{Kothalawala2024}
V.~N. Kothalawala, T.~Guruswamy, O.~Quaranta, U.~M. Patel, L.~Gades, K.~Taddei,
  A.~Yakovenko, M.~Zheng, K.~Morgan, J.~Weber, D.~Yan, D.~Swetz, I.~Makkonen,
  H.~K. Yeddu, A.~Bansil, U.~Ruett, A.~Miceli, J.~Nokelainen, and
  B.~Barbiellini, ``Extracting the electronic structure of light elements in
  bulk materials through a {{Compton}} scattering method in the readily
  accessible hard {X-Ray} regime,'' \emph{Applied Physics Letters}, vol. 124,
  no.~22, p. 223501, 2024.

\bibitem{Patel2022}
U.~Patel, T.~Guruswamy, A.~J. Krzysko, H.~Charalambous, L.~Gades, K.~Wiaderek,
  O.~Quaranta, Y.~Ren, A.~Yakovenko, U.~Ruett, and A.~Miceli, ``High-resolution
  {{Compton}} spectroscopy using {X-Ray} microcalorimeters,'' \emph{Review of
  Scientific Instruments}, vol.~93, no.~11, p. 113105, 2022.

\bibitem{Guruswamy2020}
T.~Guruswamy, L.~M. Gades, A.~Miceli, U.~M. Patel, J.~T. Weizeorick, and
  O.~Quaranta, ``Hard {{X-ray}} fluorescence measurements with {{TESs}} at the
  {{Advanced Photon Source}},'' in \emph{Journal of {{Physics}}: {{Conference
  Series}}}, vol. 1559.\hskip 1em plus 0.5em minus 0.4em\relax IOP Publishing,
  2020, p. 012018.

\bibitem{Guruswamy2021}
T.~Guruswamy, L.~Gades, A.~Miceli, U.~Patel, and O.~Quaranta, ``Beamline
  {{Spectroscopy}} of {{Integrated Circuits With Hard X-Ray Transition Edge
  Sensors}} at the {{Advanced Photon Source}},'' \emph{IEEE Transactions on
  Applied Superconductivity}, vol.~31, no.~5, pp. 1--5, 2021.

\bibitem{Lee2019}
S.-J. Lee, C.~J. Titus, R.~Alonso~Mori, M.~L. Baker, D.~A. Bennett, H.-M. Cho,
  W.~B. Doriese, J.~W. Fowler, K.~J. Gaffney, A.~Gallo, J.~D. Gard, G.~C.
  Hilton, H.~Jang, Y.~I. Joe, C.~J. Kenney, J.~Knight, T.~Kroll, J.-S. Lee,
  D.~Li, D.~Lu, R.~Marks, M.~P. Minitti, K.~M. Morgan, H.~Ogasawara, G.~C.
  O’Neil, C.~D. Reintsema, D.~R. Schmidt, D.~Sokaras, J.~N. Ullom, T.-C.
  Weng, C.~Williams, B.~A. Young, D.~S. Swetz, K.~D. Irwin, and D.~Nordlund,
  ``Soft {{X-ray}} spectroscopy with transition-edge sensors at {{Stanford
  Synchrotron Radiation Lightsource}} beamline 10-1,'' \emph{Review of
  Scientific Instruments}, vol.~90, no.~11, p. 113101, 2019.

\bibitem{formfactor-web}
\BIBentryALTinterwordspacing
{{FormFactor Home}}. FormFactor, Inc. [Online]. Available:
  \url{https://www.formfactor.com/}
\BIBentrySTDinterwordspacing

\bibitem{mass_github}
\BIBentryALTinterwordspacing
J.~Fowler and G.~O'Neil, ``{Microcalorimeter Analysis Software System
  (MASS)},'' National Institute of Standards and Technology, Nov. 2024.
  [Online]. Available: \url{https://github.com/usnistgov/mass}
\BIBentrySTDinterwordspacing

\bibitem{Fowler2016}
J.~W. Fowler, B.~K. Alpert, W.~B. Doriese, Y.~I. Joe, G.~C. O'Neil, J.~N.
  Ullom, and D.~S. Swetz, ``The practice of pulse processing,'' \emph{Journal
  of Low Temperature Physics}, vol. 184, no.~1, pp. 374--381, 2016.

\bibitem{Amorim1988}
P.~Amorim, L.~Salgueiro, F.~Parente, and J.~G. Ferreira, ``Widths of some {{L
  X-ray}} lines of iridium, platinum, gold, thorium and uranium,''
  \emph{Journal of Physics B: Atomic, Molecular and Optical Physics}, vol.~21,
  no.~23, pp. 3851--3856, 1988.

\bibitem{Irwin2005}
K.~Irwin and G.~Hilton, ``Transition-edge sensors,'' in \emph{Cryogenic
  Particle Detection}.\hskip 1em plus 0.5em minus 0.4em\relax Springer, Berlin,
  Heidelberg, 2005, pp. 63--150.

\bibitem{Cucchetti2024}
E.~Cucchetti, S.~J. Smith, M.~C. Witthoeft, M.~Eckart, F.~Pajot, P.~Peille, and
  F.~S. Porter, ``Advanced {{Energy Scale Correction Techniques}} for the
  {{X-ray Transition Edge Sensors}} of the {{Athena}} mission,'' \emph{Journal
  of Low Temperature Physics}, vol. 216, no.~1, pp. 292--301, 2024.

\bibitem{Porter2016}
F.~S. Porter, M.~P. Chiao, M.~E. Eckart, R.~Fujimoto, Y.~Ishisaki, R.~L.
  Kelley, C.~A. Kilbourne, M.~A. Leutenegger, D.~McCammon, K.~Mitsuda,
  M.~Sawada, A.~E. Szymkowiak, Y.~Takei, M.~Tashiro, M.~Tsujimoto, T.~Watanabe,
  and S.~Yamada, ``Temporal {{Gain Correction}} for {{X-ray Calorimeter
  Spectrometers}},'' \emph{Journal of Low Temperature Physics}, vol. 184,
  no.~1, pp. 498--504, 2016.

\bibitem{Yan2022}
D.~Yan, J.~C. Weber, T.~Guruswamy, K.~M. Morgan, G.~C. O’Neil, A.~L. Wessels,
  D.~A. Bennett, C.~G. Pappas, J.~A. Mates, J.~D. Gard, D.~T. Becker, J.~W.
  Fowler, D.~S. Swetz, D.~R. Schmidt, J.~N. Ullom, T.~Okumura, T.~Isobe,
  T.~Azuma, S.~Okada, S.~Yamada, T.~Hashimoto, O.~Quaranta, A.~Miceli, L.~M.
  Gades, U.~M. Patel, N.~Paul, G.~Bian, and P.~Indelicato, ``Absolute {{Energy
  Measurements}} with {{Superconducting Transition-Edge Sensors}} for {{Muonic
  X-ray Spectroscopy}} at 44~{{keV}},'' \emph{Journal of Low Temperature
  Physics}, vol. 209, no.~3, pp. 271--277, 2022.

\bibitem{nist_xraydb}
\BIBentryALTinterwordspacing
R.~Deslattes, E.~K. Jr., P.~Indelicato, L.~de~Billy, E.~Lindroth, J.~Anton,
  J.~Coursey, D.~Schwab, C.~Chang, R.~Sukumar, K.~Olsen, and R.~Dragoset,
  ``{X-ray} transition energies (version 1.2),'' National Institute of
  Standards and Technology, Gaithersburg, MD, Tech. Rep., 2005. [Online].
  Available: \url{http://physics.nist.gov/XrayTrans}
\BIBentrySTDinterwordspacing

\bibitem{Fowler2017}
J.~W. Fowler, B.~K. Alpert, D.~A. Bennett, W.~B. Doriese, J.~D. Gard, G.~C.
  Hilton, L.~T. Hudson, Y.-I. Joe, K.~M. Morgan, G.~C. O'Neil, C.~D. Reintsema,
  D.~R. Schmidt, D.~S. Swetz, C.~I. Szabo, and J.~N. Ullom, ``A reassessment of
  absolute energies of the {X-Ray} {{L}} lines of lanthanide metals,''
  \emph{Metrologia}, vol.~54, no.~4, pp. 494--511, 2017.

\bibitem{Fowler2022}
J.~W. Fowler, B.~K. Alpert, G.~C. O'Neil, D.~S. Swetz, and J.~N. Ullom,
  ``Energy {{Calibration}} of {{Nonlinear Microcalorimeters}} with
  {{Uncertainty Estimates}} from {{Gaussian Process Regression}},''
  \emph{Journal of Low Temperature Physics}, vol. 209, no.~5, pp. 1047--1054,
  Dec. 2022.

\end{thebibliography}

\end{document}